\documentclass[10pt,conference]{IEEEtran}

\IEEEoverridecommandlockouts

\usepackage{graphicx}
\usepackage{caption}
\usepackage{subcaption}
\usepackage{paralist}
\usepackage[compact]{titlesec}
\titlespacing{\section}{0pt}{0.1ex}{0ex}
\titlespacing{\subsection}{0pt}{0.1ex}{0ex}
\titlespacing{\subsubsection}{0pt}{0.1ex}{0ex}    

\begin{document}
%
\title{PHOENI2X - A European Cyber Resilience Framework With Artificial-Intelligence-Assisted Orchestration, Automation and Response Capabilities for Business Continuity and Recovery, Incident Response, and Information Exchange}

\author{\\[-12ex]\IEEEauthorblockN{Konstantinos Fysarakis\IEEEauthorrefmark{1},
Alexios Lekidis\IEEEauthorrefmark{2},
Vasileios Mavroeidis\IEEEauthorrefmark{3},
Konstantinos Lampropoulos\IEEEauthorrefmark{4},
George Lyberopoulos\IEEEauthorrefmark{5},\\ 
Ignasi Garcia-Milà Vidal\IEEEauthorrefmark{6},
José Carles Terés i Casals\IEEEauthorrefmark{7},
Eva Rodriguez Luna\IEEEauthorrefmark{8}, 
Alejandro Antonio Moreno Sancho\IEEEauthorrefmark{9},\\
Antonios Mavrelos\IEEEauthorrefmark{10}, 
Marinos Tsantekidis\IEEEauthorrefmark{11},
Sebastian Pape\IEEEauthorrefmark{12}, 
Argyro Chatzopoulou\IEEEauthorrefmark{13},
Christina Nanou\IEEEauthorrefmark{14},\\
George Drivas\IEEEauthorrefmark{15},
Vangelis Photiou\IEEEauthorrefmark{16},
George Spanoudakis\IEEEauthorrefmark{1}, and
Odysseas Koufopavlou\IEEEauthorrefmark{4}\IEEEoverridecommandlockouts}

\IEEEauthorrefmark{1} Sphynx Analytics Limited, \IEEEauthorrefmark{2} Public Power Corporation S.A., Greece, \\ \IEEEauthorrefmark{3} University of Oslo, Norway, \IEEEauthorrefmark{4} University of Patras, Greece, \\ \IEEEauthorrefmark{5} Cosmote Kinites Tilepikoinonies A.E., Greece, \IEEEauthorrefmark{6} Worldsensing, Spain, \\ \IEEEauthorrefmark{7} Ferrocarriles de la Generalitat, Spain, \IEEEauthorrefmark{8} Universitat Politecnica de Catalunya, Spain, \\ \IEEEauthorrefmark{9} Atos IT Solutions and Services Iberia S.L., Spain, \IEEEauthorrefmark{10} Nodalpoint Systems, Greece, \\ \IEEEauthorrefmark{11} AEGIS IT Research AG, Germany, \IEEEauthorrefmark{12} Social Engineering Academy, Germany, \\ \IEEEauthorrefmark{13} APIROPLUS Solutions Ltd., Cyprus, \IEEEauthorrefmark{14} EUNOMIA Limited, Ireland, \\ \IEEEauthorrefmark{15} Ministry Of Digital Governance, Greece, \IEEEauthorrefmark{16} Archi Psifiakis Asfaleias, Cyprus \\
\footnotesize{Email:\IEEEauthorrefmark{1}fysarakis@sphynx.ch}\vspace{-2ex}}


\maketitle

\begin{abstract}
As digital technologies become more pervasive in society and the economy, cybersecurity incidents become more frequent and impactful. According to the NIS and NIS2 Directives, EU Member States and their Operators of Essential Services must establish a minimum baseline set of cybersecurity capabilities and engage in cross-border coordination and cooperation. However, this is only a small step towards European cyber resilience. In this landscape, preparedness, shared situational awareness, and coordinated incident response are essential for effective cyber crisis management and resilience. Motivated by the above, this paper presents PHOENI2X, an EU-funded project aiming to design, develop, and deliver a Cyber Resilience Framework providing Artificial-Intelligence-assisted orchestration, automation and response capabilities for business continuity and recovery, incident response, and information exchange, tailored to the needs of Operators of Essential Services and the EU Member State authorities entrusted with cybersecurity.\end{abstract}


%
\IEEEpeerreviewmaketitle

\section{Introduction} \label{sec:intro}
As our societies and economies increasingly rely on digital infrastructures, cybersecurity incidents become more frequent, diversified, and impactful. In fact, ENISA’s recent Threat Landscape reports (e.g., for 2021 and 2022\footnote{https://www.enisa.europa.eu/publications/enisa-threat-landscape-2022}) highlight that cybersecurity risks are increasingly becoming harder to assess and interpret, due to the growing complexity of the threat landscape, the adversarial ecosystem and the expansion of attack surfaces. Hence, threats in cyberspace endanger the European long-term objectives (e.g., the Digital Single Market that aims to enhance Europe's position as a world leader in the digital economy).

The introduction of the NIS Directive (2016/1148) \cite{markopoulou2019new}, pushed EU Member States (MS) and their Operators of Essential Services (OES) to achieve a minimum baseline set of cybersecurity capabilities and engage in cross-border coordination and cooperation. The introduction of NIS2, an update of NIS, increases the number of sectors classified as OES from 19 to 35, imposes stricter requirements upon them, and consequently affects other third parties that must further reinforce this strategic push for increased cybersecurity and resilience. These, along with several additional European initiatives (e.g., the European Cyber Defence Policy and the European Cyber Resilience Act), are underway to strategically address this challenge, essentially implementing the EU’s cybersecurity Strategy. These efforts and the associated entities that will be created to support their implementation (e.g., Joint Cyber Unit) are necessary for increasing the collective resilience, incident detection, and defence capabilities, as well as the operational and technical coordination of EU countries for crisis management \cite{samtani2017exploring}.

In this landscape, preparedness \cite{deb2018predicting}, shared situational awareness \cite{schauer2017adaptive}, and coordinated Incident Response (IR) \cite{souppaya2017guide} are essential, not just for effective crisis management and cyber resilience, but also for driving strategic/political decisions that will effectively tackle threats that threaten the well-being of the EU. This is a challenging activity that requires the development of new services and the enhancement of existing ones with automated procedures \cite{cyberSec}. PHOENI2X\footnote{https://cordis.europa.eu/project/id/101070586} is a Horizon Europe project that aims to address this challenge by providing tools and mechanisms as essential enablers to ensure cyber resilience.

The concrete contributions of PHOENI2X presented in this paper are (a) automation/orchestration mechanisms for business continuity and IR; (b) actionable, relevant, and timely Cyber Threat Intelligence (CTI) for increased threat situational awareness, interoperable standardised alerting, reporting, and information exchange mechanisms, and access to early warning systems; (c) structured training through realistic Cyber Ranges (CR) for preparedness.

The rest of the paper is organized as follows. Section \ref{sec:framework} presents the concept and approach of the PHOENI2X project, which aims to design, develop, and deliver a Cyber Resilience Framework (CRF) providing AI-assisted orchestration, automation and response capabilities, covering business continuity, incident response, and information exchange, tailored to the needs of OES and of the EU MS National Authorities entrusted with cybersecurity. Section \ref{sec:innovation} presents the main innovation areas and relevant research challenges. Then, Section \ref{sec:useCases} demonstrates three characteristic OES-focused use cases, covering energy, transport and healthcare, which are used to validate the proposed approach. Finally, Section \ref{sec:conclusion} provides the concluding remarks and next steps.


\section{The PHOENI2X Cyber Resilience Framework} \label{sec:framework}

At the core of the PHOENI2X concept is the deployment of PHOENI2X Cyber Resilience Centres (PHOENI2X CRCs) at OES premises, adopting a recently-published conceptual blueprint in support of architecting and establishing interoperable Cyber Security Operations Centres that combine capacity for Shared Situational Awareness, Coordinated Response, and Joint Preparedness \cite{Blueprint}. More specifically, the CRCs will integrate capacities for:

\noindent - Enhanced \textit{Situational Awareness} with AI-assisted Prediction, Prevention, Detection and Response capabilities, and Business Risk Impact Assessment-based Prioritisation.

\noindent - Proactive and reactive \textit{Resilience Automation, Orchestration, and Response (ROAR)} mechanisms, providing Business Continuity, Recovery and Cyber and Physical IR.

\noindent - Increased \textit{Preparedness} through relevant Serious Games and realistic Resilience Cyber Range (RCR) Assessment and Training.

\noindent - Timely \textit{Information Exchange} between OES, National Authorities and EU actors, leveraging interoperable and standardised alerting/reporting mechanisms and processes.

\begin{figure*}[bthp!]
\vspace{-4ex}
     \centering
         \includegraphics[scale=.56]{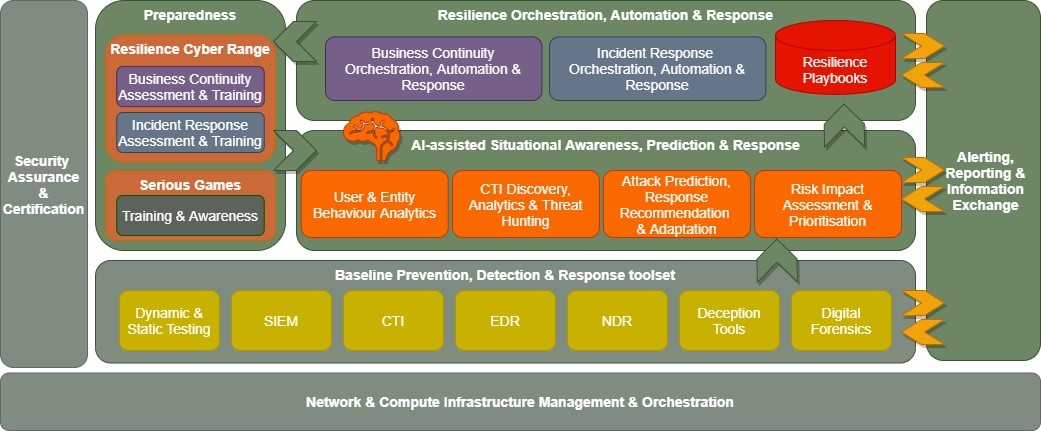}
     \caption{PHOENI2X Conceptual Architecture}
     \label{fig:framework}
     \vspace{-6ex}
\end{figure*}

A high-level architectural view of the PHOENI2X CRCs is provided in Fig. \ref{fig:framework}, while the key building blocks are detailed in the subsections that follow.

\subsection{Baseline prevention, detection and response toolset}

To provide the baseline prevention, detection, and response capabilities that the rest of the framework will build upon, PHOENI2X integrates a toolset comprising open-source tools, covering: Dynamic and static vulnerability testing (e.g., OpenVAS\footnote{https://www.openvas.org/}); Security Information and Event Management (SIEM; e.g., Wazuh\footnote{https://wazuh.com/}), CTI management and sharing (e.g., OpenCTI\footnote{https://www.filigran.io/en/products/opencti/}); Endpoint Detection and Response (EDR; e.g., OSSEC\footnote{https://www.ossec.net/}); Network Detection and Response (NDR; e.g., Suricata\footnote{https://suricata.io/}); Deception tools (e.g., Dionaea\footnote{https://github.com/DinoTools/dionaea}), and; Forensics tools (e.g., GRR\footnote{https://github.com/google/grr}).

\subsection{AI-assisted situational awareness, prediction and response}

PHOENI2X will integrate and combine different AI technologies to assist its Situational Awareness, Prediction and Response capabilities, including: 

\noindent - \textit{User and Entity Behaviour Analytics}, focusing on the use of ML-based analytics for the creation of a behavioural baseline for each user and entity (e.g., systems, devices), using any deviations to provide early warnings for security events.

\noindent - \textit{CTI Discovery, Analytics and Threat Hunting}, leveraging knowledge graph technology to connect and correlate heterogeneous data, perform intelligible queries for analytics and threat hunting, and use reasoning to discover new information (inference) in near real-time. In addition, the use of NLP will assist with extracting relevant artefacts from reports and other sources, such as from the dark web, and will be incorporated into the knowledge graph. The PHOENI2X knowledge graph technology will communicate seamlessly with CTI platforms like MISP and OpenCTI. A baseline STIX 2.1 ontology, also known as the Threat Actor Context Ontology (TAC ontology\footnote{https://www.oasis-open.org/committees/tac}), will be used and extended based on the infrastructures and use cases of the end users.

\noindent - \textit{Attack Prediction, Response Recommendation and Adaptation}, encompassing attack categorization to identify the scope and impact of potential attacks, attack prediction through training ad-hoc behavioural models for each system, and attack response and adaptation, including proactive actions and mitigation strategies.

\noindent - \textit{Risk Impact Assessment and Prioritisation}, featuring a near real-time risk assessment engine for the OES (through pre-defined machine-readable model rules: a qualitative model based on DEXi\footnote{https://kt.ijs.si/MarkoBohanec/dexi.html} and a quantitative one to be developed within PHOENI2X), thus facilitating response prioritisation.

\subsection{Resilience orchestration, automation and response (ROAR)}
Fundamental to the PHOENI2X Resilience Orchestration and Response capabilities are the Resilience Playbooks (RPs). RPs will provide a structured, machine-processable encoding of a sequence of actions comprising the organisation’s business continuity, recovery, and IR processes. Each action represents a fundamental activity (e.g., adding a rule to a firewall). Thus, through RPs, organisations will be able to specify, automate the execution (via the purpose-built execution and orchestration engine), monitor the progress, and assess the effectiveness of all their business continuity, recovery, and IR -related processes. To achieve the above, RPs in PHOENI2X will adopt and extend the recently released OASIS Collaborative Automated Course of Action Operations (CACAO) specification\footnote{https://docs.oasis-open.org/cacao/security-playbooks/v1.0/security-playbooks-v1.0.html}.

\subsection{Preparedness}
Towards increased Preparedness, PHOENI2X will provide realistic scenario assessment and training capabilities through the integration of an RCR and Serious Games.
The RCR will support OES preparedness in two ways: (i) it will enable the assessment of defined RPs of all types in a realistic environment that emulates and simulates, as needed, the involved cyber systems and their interactions (e.g., to find gaps or inefficiencies, adapting them as needed); (ii) it will provide hands-on training to OES staff in the business continuity, recovery and IR procedures encoded in the RPs (e.g., to ensure everyone is accustomed to a new BC approach). 


Serious Games will be used for training that specifically targets the human factor (many times considered the weakest link in security). The serious gaming capability will support the RCR, focusing on the Training and Awareness of employees on different cyber attacks, threat elicitation, and improving organizational defences that depend critically on human factors (e.g., social engineering and phishing attacks). 

\subsection{Alerting, reporting and information exchange}

This subsystem will comprise an information exchange platform to address the common need for standardized and coordinated cybersecurity notifications both in everyday operations and, most importantly, when a significant cyber incident is manifested. 
In addition to the above reporting mechanism, the subsystem will be used to support the exchange of (i) technical-level indicators from the Baseline Prevention, Detection and Response toolset; (ii) AI-generated insights, early-warning alerts, contextualised information - CTI, models, and other shareable information from the AI-assisted Situational Awareness, Prediction and Response enablers; (iii) RPs from the ROAR subsystem; (iv) training programmes from the RCR. An overarching objective when developing the above will be the provision of a coordinated response-driven system that aims to, eventually, link all the EU-level relevant parties at the strategic and political level, as well as cybersecurity IR actors (CSIRTs network, ENISA, NIS CG, CyCLONe) and relevant third parties (e.g., private entities). 

\subsection{Security assurance and certification}
Since the introduction of PHOENI2X in the OES environment may increase its attack surface, PHOENI2X CRCs will include a continuous, evidence-based Security Assurance and Certification solution. This will provide comprehensive assessment coverage of every component of the PHOENI2X CRCs and their operation, covering all assets (e.g., network, compute, data, processes) comprising each CRC.

\subsection{Network infrastructure management and orchestration} 
PHOENI2X will use cloud resources and orchestration tools to deploy and manage its framework and support the adaptation actions of the OES infrastructure encoded within RPs. To this end, a cloud platform will be able to host core network components and support Network Function Visualisation and Multi-access Edge Computing deployments.

\section{Key Innovation Areas and Research Challenges} \label{sec:innovation}

\subsection{ML-based user and entity behaviour analytics (UEBA)}

UEBA is an advancement of User Behaviour Analytics, using big data to model and identify normal and abnormal behaviour of machines and humans. The introduced term ‘entity’ is used to describe the ability to model behaviours of critical infrastructures (CI), IT systems, organisations, etc. By using unsupervised machine learning algorithms, UEBA detects changes in the behaviour of the communication patterns between servers and endpoints \cite{salitin2018role} to detect possible attacks. Apart from external attacks, UEBA is also used for detecting insider threats and attacks. Having set a baseline user profile, anomalous activity can be detected by calculating the deviations from the normal behaviour \cite{khaliq2020role}. In addition, by combining UEBA with data visualization or Long Sort Term Memory (LSTM) neural networks, the use cases can expand to ransomware detection \cite{ganfure2020deepguard} and anomaly detection \cite{rengarajan2021anomaly}.

In PHOENI2X, an AutoML-based system will be adopted to create optimal classification, regression, and statistic models for UEBA use cases by creating pipelines for pre-processing and validating big data while also integrating AI explainability features, to support transparency and audit towards trustworthy AI. Innovation will also come with the integration of human behaviour aspects that are currently not considered in the literature. More specifically, in Social Psychology, the concept of Persuasion refers to an active attempt to change a person’s mind \cite{petty2018attitudes}. Therefore, persuasion is an essential element of a social engineering attack, which is often part of the initial step of Tactics, Techniques and Procedures (TTPs) of given Advanced Persistence Threats (APT) groups. Generic patterns for adaptation in UEBA use cases should be created that identify potential persuasion attempts. These can be supported and reinforced through the ingestion of training (e.g., RCR and Serious Games) results.

\subsection{AI-assisted categorisation, prediction and response} 

The impact of an attack will severely be affected by the reaction time of defenders to initiate a response and the time required to successfully complete the process. It seems reasonable to state that the sooner a response is initiated during an attack, the smaller the impact on the organisation, which pushes defenders to incorporate proactive defence systems based on some attack-predictive strategies. In this context, Deep-Learning-based approaches are widely used for attack prediction. For instance, Generative Adversarial Networks (GANs) are used to address the dynamic change of threats \cite{vu2020deep} and leverage Transfer Learning techniques to minimize the inherent computational cost \cite{zhao2019transfer}, proposing pre-trained language models intended to predict how exploitable the system’s vulnerabilities are \cite{yin2020apply}. Moreover, the utilisation of AI and data mining to predict cybersecurity incidents and produce system maintenance actions to avoid attack realization has been investigated \cite{sun2018data}. To define and optimize the set of such maintenance actions, it is a must to assess the expected attack(s) impact on the system. To that end, beyond the vulnerability and impact analysis traditionally used, attack categorization is fundamental. Indeed, attack categorization is a key pillar in any cybersecurity framework. This becomes imperative when aiming to avoid system impacts that may disrupt business processes. A relevant model for attack pattern categorization is the MITRE ATT\&CK Framework\footnote{https://attack.mitre.org}, which identifies TTPs used by attackers. This facilitates the categorization of attacks and consequently helps to establish their potential impact.

An Attack Categorisation Modelling (ATM) tool proposed in PHOENI2X aims to address this critical aspect by analysing the system behaviour and, based on data analytics, identifying the real effects a potential attack may have on the system. The proposed technology complements MITRE ATT\&CK by adding attack-modelled profiles along with predictive impacts that may help cover the continuous evolution of the attacks' nature. The ATM tool will provide a data repository that may cover a broad spectrum of attacks and scenarios and will aim at producing a Predictive Maintenance (PMEM) tool that is continuously fed with data coming from past experiences through a data sharing model and turns out into a set of specific actions customised for each sector and scenario. In short, PMEM uses supervised and unsupervised learning approaches to predict intrusions and propose specific actions as proactive actions to prevent the attack or as a response to the attack.

\subsection{ML-based discovery, extraction, and CTI analysis}
The continuously evolving CTI landscape requires the mobilization of all available sources and evidence, including information in articles, blogs, hacker forums, malicious assets, and other forms of human-readable text \cite{souppaya2017guide}. Automatic extraction to machine-readable forms is a challenging task that involves NLP, ML techniques and data mining. Existing  background work on ML and NLP-based extraction of CTI \cite{sun2021automatic} is present. However, NLP solely lacks to accomplish this task given limitations such as the domain specificity and the complexity of semantic analysis \cite{samtani2017exploring} \cite{deb2018predicting}. Social network analysis \cite{biswas2018leadership} can also be used towards this goal, along with sentiment analysis \cite{li2014identifying}, and Convolutional Neural Networks (CNNs); either to measure the CTI quality \cite{schlette2021measuring}, extract data from real hacker forums or malicious assets (crypters, keyloggers, web and database exploits) \cite{samtani2017exploring}, apply classification through CNNs \cite{deliu2017extracting} or identify key hackers for keylogging tools \cite{biswas2018leadership}. Sentiment analysis has been applied to hacker forum posts to predict cyber events \cite{deb2018predicting}, or examine hacker behaviour in dark forums and apply classification through a logistic regression model. Other studies focus on proactive CTI by analysing the source code in hacker forums via Deep Transfer Learning for Exploit Labelling (DTL-EL) \cite{ampel2020labeling} or extracting indicators of compromise (IOC) through word embedding in CNNs.

In PHOENI2X, the above will be extended by defining an AutoML-based analysis pipeline that will gather, via NLP and data mining, pertinent data from diverse sources (e.g., hacker forum texts, scraping surface web or dark web, cyber event attacks with ground truth data from public organisations, social media channels). This will be followed by sentiment analysis and contextualisation of the extracted data. There are several directions that will be explored for each type of data, leveraging the flexibility of the AutoML systems, including supervised sentiment analysis (with labelled training data), unsupervised, lexical-based methods, deep learning, and dictionary-based sentiment algorithm. Available tools for sentiment analysis will also be investigated, including VADER \cite{hutto2014vader}, LIWC \cite{pennebaker2001linguistic}, and SentiStrength \cite{thelwall2017heart} as well as several alternatives for deep learning (Classic, Convolutional, Recurrent, GANs, Self-Organizing Maps, etc.).

\subsection{CTI - contextualization and threat hunting}

With the increasing number and complexity of cyber attacks, organisations have seen the need to share CTI, and the practice has grown significantly. An ENISA report on the opportunities and limitations of current Threat Intelligence Platforms (TIP) \cite{beard2017exploring} provides an overview of some of the main TIPs (including e.g. CRITs, CIF, MISP, OTX or ThreatExchange) and includes among the limitations in the current state and usage of TIPs the following ones: (i) the data shared is “too voluminous and complex to be actioned”; (ii) there is “limited technology enablement in threat triage and relevancy determination”; (iii) there are “trust-related issues”, and; (iv) “qualities of shared threat data and TIP limitations” (indicating the need of including context in shared data). Tounsi et al. \cite{tounsi2018survey} also highlight the need to filter the information collected and put focus on their own internal vulnerabilities and weaknesses to prevent cyber attacks. Several EU-funded projects try to address some of these limitations. In the DiSIEM\footnote{https://disiem.lasige.di.fc.ul.pt/}, an extension of the quality assessment processes to improve the information sharing capabilities of TIPs complementing static information about the monitored infrastructures with real-time CTI to enrich information received from OSINT is proposed \cite{faiella2019enriching}. In CyberSec4Europe\footnote{https://cybersec4europe.eu/} different assets have been developed for increasing trustworthiness, quality and reliability in CTI sharing \cite{cyberSec}. In FINSEC\footnote{https://www.finsec-project.eu/}, the FINSTIX Data Model is proposed as a standard-based approach to model and represent CTI for predictive and collaborative security of Financial Infrastructures.

A key challenge to be addressed is to operate upon incoming CTI indicators, including those discovered and extracted from heterogeneous channels, contextualizing and qualifying them with static information about the infrastructure but also with real-time alerts generated by the underlying baseline toolset in order to determine with a score its relevance and actionability in the organisation. The enriched data can be sent for its processing and visualization to other tools, such as the SIEM, while also informing the AI-assisted response recommendation and prioritisation tools.

\subsection{Risk assessment-based alert and response prioritization} 
To assess risk (a key requirement for all organisations), most methodologies follow the traditional approach where the likelihood of occurrence of a security incident and its potential impact on the system or in the organisation are considered to define a qualitative risk scale. The Factor Analysis of Information Risk model can also be used in conjunction with the above to evaluate factors and establish probabilities that achieve more accurate risk-based quantification\footnote{https://www.fairinstitute.org/what-is-fair}. Three main traditional approaches to risk modelling are found in the literature: tree-based notations (e.g., event tree analysis \cite{iec201062502}), graph-based notations (e.g., the model-based risk analysis approach CORAS \cite{lund2011risk}, and table-based techniques \cite{ishtiaque2019hazard}). Attempts to improve the situational awareness in cybersecurity risk assessment can be found for specific sectors (e.g., the H2020 project MITIGATE presents a collaborative evidence-driven Maritime Supply Chain Risk Assessment approach, g-MSRA \cite{schauer2017adaptive}, that analyses potential threats to the maritime supply chains considering information coming from Open-Source Intelligence). 

There is a need to go beyond the State-of-the-Art (SotA) by improving the proactive and reactive response to risks by offering a qualitative and quantitative business risk assessment together with a list of potential mitigation measures and countermeasures classified in order of priority. They can include preventive recommendations associated with the risks identified but also manually added by the security analysts. 

\subsection{Security automation, orchestration and response} 
The performance of IR systems and processes within an organisation can be assessed by quantitative metrics such as the mean time to acknowledge and the mean time to remediate. To improve such metrics, Security Orchestration, Automation and Response (SOAR) and Extended Detection and Response (XDR) offerings that utilize IR playbooks are currently emerging. In their basic form, IR playbooks are structured documents with all the necessary steps for attack detection and mitigation. Such playbooks and playbook compilation guides are provided by many security organisations and consortia such as NIST \cite{souppaya2017guide}, SANS\footnote{https://www.sans.org/presentations/ir-playbooks/}, governments\footnote{https://github.com/cisagov/shareable-soar-workflows} and other parties\footnote{https://www.incidentresponse.com/playbooks/}. Nowadays, the number of SOAR systems that utilize machine-readable playbooks is constantly increasing. These advanced playbooks are not just a structured rule set or guidelines but incorporate executable workflows enabling automation. In most cases, executable playbooks are not publicly available and are shipped with the proprietary SOAR and XDR systems that utilize them\footnote{https://docs.fortinet.com/document/fortisoar/6.0.0/playbooks-guide}. Another major advantage of executable playbooks and the systems that support them is their graphical representation, usually as a graph, and the ability to edit and fine-tune them in a user-friendly manner. As the need for publicly available, executable, extendable and shareable playbooks is nowadays apparent, OASIS has recently developed the CACAO specification for such playbooks. The aim of such endeavours is to eventually establish a common specification for compiling IR playbooks able to be shared across various sectors and organisations, executable by a plethora of SOAR and XDR systems. In addition, works on the advantages of coupling and exchanging playbooks with CTI and prototypical implementations of such mechanisms have emerged \cite{mavroeidis2021integration, mavroeidis2022cybersecurity}.


While playbooks begin to rise in popularity, the cybersecurity IR landscape is currently dominated by proprietary playbook specifications or BPML-based high-level playbooks that lack any orchestration and execution capabilities. PHOENI2X aims to go beyond the current SotA in the domain by offering: (i) playbook execution and orchestration capabilities that go beyond the current state-of-the-art academic and commercial solutions; (ii) the first, at the time of writing, definition of executable business continuity playbooks focusing on business processes and maintaining required service levels; (iii) the first CACAO-based orchestration, automation and response engine and orchestrator; (iv) novel extensions to CACAO to support business continuity playbooks and other advanced features of the platform; (v) specification of novel what-if analysis scenario playbooks, e.g., involving hypothetical components and tools that an organisation may be considering to introduce; (vi) tight integration with a CR to assess playbooks in a realistic simulated/emulated environment and to also derive training from said playbooks; (vii) inclusion of machine-speed information exchange mechanisms within the playbooks, including playbook sharing but also alerting and generating reports aligned with the OES requirements.

\subsection{Cyber range training}
CRs as means of training and employee assessment are becoming increasingly popular because they address the most problematic part of any cyber system, the human, by offering dynamic, hands-on, scalable and disposable environments to learn and test security concepts and emerging technologies. They are typically based on emulation (e.g., OpenStack) and simulation technologies (e.g., NS-3), and as these technologies mature, so do the CRs, offering greater fidelity and variety \cite{ukwandu2020review} on the implemented training scenarios. Especially in cybersecurity, CRs are becoming prominent solutions for training employees of different levels of expertise, from blue teams to non-technical employees (e.g., for security fundamentals and awareness training) \cite{thomas2018individual}. CRs usually offer a baseline set of features, such as predefined scenarios based on existing threats, physical or remote access to the training environment, team training, and data monitoring with tools such as OSSIM. A few offer more advanced features, such as attack simulation \cite{hildebrand2019clusus} or user experience features like drag and drop scenario specification. Research\footnote{https://www.cyberbit.com/wp-content/uploads/2016/09/CB-TnS-Print.pdf} also focuses on adaptation procedures for training scenarios based on user performance and other security indicators (e.g., penetration testing findings). Another line of research focuses on the assessment and accurate evaluation of trainees in terms of their performance on the virtual infrastructure (e.g., through evaluating user actions via regular expressions \cite{thompson2021labtainers}, or through directed graphs to model and keep track of the users' performance \cite{andreolini2020framework}).


PHOENI2X aims to go beyond the current SotA by integrating a Resilience CR, which will be uniquely able to ingest business continuity, recovery, and IR orchestrations, as encoded in the associated RPs, and automatically generate the (emulated and simulated) environments to execute them. This capability will be leveraged to assess the effectiveness and efficacy of the orchestrations encoded within said RPs, also including the what-if analyses of orchestrations involving hypothetical assets and business processes to inform decision-making. Moreover, this capability will also be used to generate hands-on training programmes to increase the preparedness of OES employees in executing the strategies encoded in the RP.

\subsection{Serious Games}
Social engineering attacks represent a continuing threat to employees of organisations. With a wide availability of different tools and information sources \cite{beckers2017structured}, it is a challenging task to keep up to date with the latest attack techniques directed to employees. A Data Breach analysis \cite{bassett2021data} reported an increase in financially motivated social engineering where the attacker directly asks for money (e.g., by impersonating CEOs or other high-level executives) or requests a purchase and transfer of online gift cards to scam employees. Additionally, adversaries may base their attacks on the latest news, such as COVID-19 ransomware \cite{saleh2020covidlock}. While certain defence methods and counteracting training methods exist \cite{schaab2017social}, at present, most of them cannot be adapted fast enough to cope with the amount and speed of variation of the attack techniques. Attempts undertaken in the past failed to achieve this goal, such as CyberCIEGE \cite{irvine2005cyberciege} and PlayingSafe \cite{newbould2009playing}.

PHOENI2X aims to go beyond the current SotA by leveraging serious games designed to raise awareness of social engineering. The games will incorporate levels that tackle different cognitive aspects and hence provide an effective learning experience. A simulation of a real-life, lesson-learned application should test the players' ability to detect attacks - an approach very similar to inoculation \cite{olanrewaju2015social}. Moreover, valuable insights will be gained by studying the effect of Serious Games integration to inform the AI-assisted situational awareness enablers and assess and support the IR and BC preparedness of OES.

\section{OES Use Cases, Requirements and Threats} \label{sec:useCases}

The PHOENI2X concept will be validated in the context of three OES-focused use cases, including: (i) an Energy use case based in Greece involving an OES (Public Power Corporation) as well as a supporting telecom provider (COSMOTE) and the National Cybersecurity Authority (NCSA, Ministry of Digital Governance) overseeing the OES; (ii) a Transport use case based in Spain involving an OES (Ferrocarrils de la Generalitat de Catalunya railway), and; (iii) a Healthcare use case based in Cyprus involving an essential solution and infrastructure provider (Nodalpoint Systems) of an OES (the General Healthcare System of Cyprus) to highlight the importance of supply chain aspects. More details on these critical use case domains and their intricacies, including the main threats identified, are provided below.

\subsection{Energy - cascading effects of cyber attacks against advanced metering infrastructure} 

\begin{figure}[bthp!]
\vspace{-2ex}
     \centering
         \includegraphics[scale=.75]{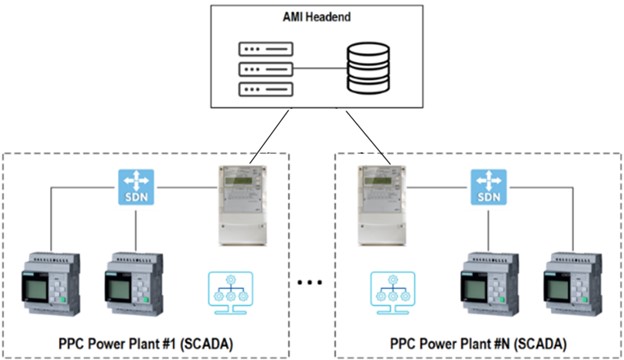}
     \caption{Typical AMI topology}
     \label{fig:amiInfra}
     \vspace{-7ex}
\end{figure}

Advanced Metering Infrastructure (AMI) is an integral part of a smart grid, consisting of technologies that allow continuous monitoring, profiling, energy auditing and, possibly, load curtailment, including both consumers and producers. The smart meter is a basic AMI component deployed on the end user premises and remotely controlled by the energy operator. Fig. \ref{fig:amiInfra} shows the interaction between power plants and the AMI Headend. Smart meters hold a significant role in the proper operation of the energy market since the retrieved measurements are used for transparent billing and market clearance as well as for determining additional actions that ensure load and supply balance. Subsequently, any disturbance in their operation, any loss or illegal access to their generated measurements, or miscommunication with the AMI Headend could result in cascading effects. Such effects are not only limited to the grid operator’s inability to accurately detect and respond to emergencies and balance the market but also to implications that can lead to market manipulation. 

A smart meter on the power plant network transmits measurements to the Headend hosted on the grid operator. In the context of this use case, PHOENI2X  will evaluate cross-sector security aspects between different types of OESs (energy and telecom). In particular, the telecom’s primary role is to provide a reliable communication channel between the individual sites (e.g., power plants) and the central AMI Headend, meeting the specific requirements of the energy operator (in terms of bandwidth, latency, jitter, etc.). To ensure high availability connectivity free of disruptions that could potentially occur in cases of attacks against the energy or telecom infrastructure, the telecom shall provide a fallback communication path (of similar capabilities/characteristics) which will become available in an automated way. Depending on the criticality of the affected operations and/or components involved, the site communication-related requirements, the application itself and the mitigation strategy of the energy operator, the telecom operator will automatically switch to the appropriate alternative technology among the available ones (e.g., 4G, 5G). 

\noindent \textbf{Requirements:} High availability requirements ($\geq$ 99.999\%); Detection and mitigation of attacks against Device Language Message Specification (DLMS) and Companion Specification for Energy Metering (COSEM).

\noindent \textbf{Key Threats:} False-data injection to forge measurements, aiming to manipulate the energy market or the Market Clearing Price (MCP); Distributed Denial of Service attacks against smart meters, aiming at availability; Attacks against the measurement database of the AMI Headend; Eavesdropping of AMI communication, resulting in unwanted energy profiling. Prominent attack scenarios also include: false data injection attack to artificially increase peak usage by an insider, thereby increasing energy demand and, possibly, MCP; Attack on the intermediary telecom infrastructure (e.g., routing attack against BGP) results in smart meter unavailability, thus, some power plants cannot participate to the load balancing market.

\subsection{Transport - cyber and physical attacks and risk management service for the railway management system}

The current railway infrastructure has been in use for a long time, dealing successfully with the operational dimension (reliability, safety, on-time performance). However, due to the vast digitisation of IT systems and infrastructure, the challenge is to add cybersecurity awareness and cyber defence countermeasures within (or operating in parallel with) the legacy system. The implementation of new ICT products (including but not limited to cloud and IoT services offered to passengers, smart ticketing, rail analytics, freight information, fleet assignment and monitoring) without properly integrating security-by-design in the railway infrastructure is bound to increase their exposure, leaving them vulnerable to a whole new set of threats and attacks.

\begin{figure}[bthp!]
    \vspace{-2ex}
     \centering
         \includegraphics[scale=.75]{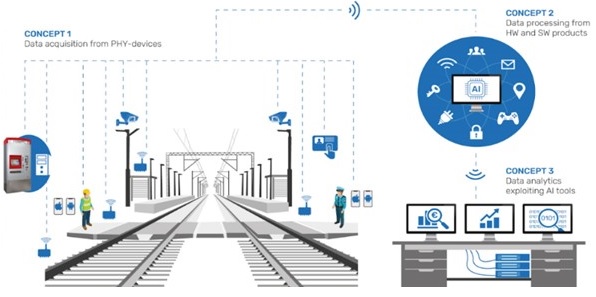}
     \caption{Railway infrastructure monitoring/management services}
     \label{fig:railway}
     \vspace{-3ex}
\end{figure}

This use case involves the provision of beyond SotA cybersecurity services (see Fig. \ref{fig:railway}) tailored to a digital railway infrastructure. These services will improve the proactive strategy to predict cyber attacks, identifying and isolating security and safety threats by dividing the system into 3-layers (data collection, processing and analytics). The experiments will cover: i) a complete set of functionalities for protection of the deployed rail infrastructure against the main physical and cybersecurity threats; ii) endpoint security for remote devices with access to the core system; iii) tools for automatic system vulnerability assessment when new products/devices/features are added to the system; iv) user behaviour monitoring solutions for protection against malicious insiders; and v) monitoring tools for continuous anomaly detection.

\noindent \textbf{Requirements:} Incident/threat detection; Sensitive data protection; Authentication and integrity; System authorization; Delay; System availability; Cost-effectiveness.

\noindent \textbf{Key Threats:} Physical and cybersecurity threats that can cause dangerous situations in railways. Physical incidents alone can have serious effects (e.g., Rheintalbahn incident in Rastatt, Germany, in 2017, closing tracks for 51 days, and only 10\% to 20\% of freight traffic could be handled via alternative routes\footnote{https://blog.src-consulting.com/en/how-wannacry-compromised-it-security-worldwide}). Prominent attack scenarios also include: External cyber/physical attacks to the rail infrastructure; Cyber attacks to the railway control room; and malicious insiders.


\subsection{Healthcare - cyber attacks aiming to cripple the public healthcare system}

\begin{figure}[bthp!]
    \vspace{-3ex}
     \centering
         \includegraphics[scale=.7]{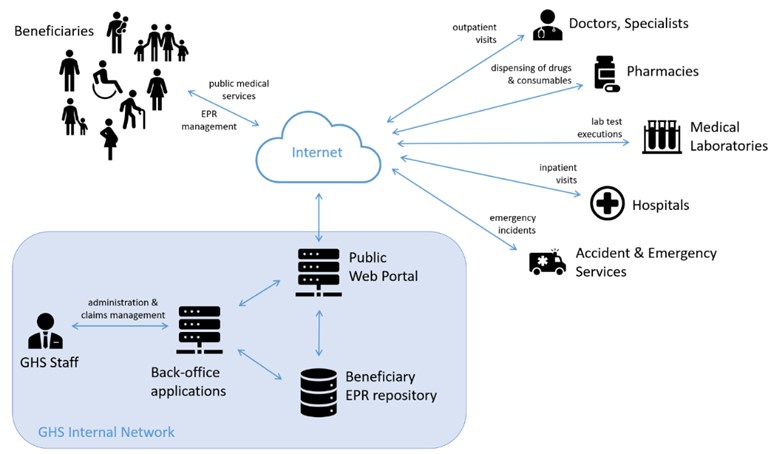}
     \caption{General Health System high-level architecture}
     \label{fig:health}
     \vspace{-6ex}
\end{figure}

A typical General Health System (GHS) infrastructure connects healthcare providers with beneficiaries to provide medical e-services to all eligible individuals (see Fig. \ref{fig:health}). For sensitivity reasons, each beneficiary can control what medical data is visible to the healthcare providers. A public website provides access to thousands of registered healthcare professionals and millions of beneficiaries, handling thousands of visits/hospitalisations and medical claims daily. Aside from the public portal, the GHS internal network consists also of a series of back-office applications, where qualified GHS staff carry out administrative functions such as enrolling beneficiaries, validating healthcare providers and maintaining catalogues of available drugs, lab tests and medical procedures. Arguably the most important function performed by the GHS staff is approving/managing claims for healthcare services. This involves reimbursing healthcare providers for goods/services offered but also checking whether claims are properly substantiated, identifying over-use of drugs or services, or detecting other fraudulent behaviour patterns. The GHS staff clears millions worth of healthcare goods/services per month. 

Typical business processes through the public web portal include doctors or hospitals registering a new beneficiary visit (inpatient or outpatient), a doctor prescribing drugs or consumables, a doctor ordering lab tests, a doctor issuing a referral to another doctor, a pharmacy dispensing drugs, a medical laboratory executing lab tests and Accident and Emergency providers handling an emergency incident. Further business processes through the public web portal include the enrolment of family members and, most importantly, viewing/managing their Personal Health Records (PHR) or Electronic Patient Records (EPR). Finally, business processes handled by the back-office applications include reimbursing healthcare providers and administering the system.

\noindent \textbf{Requirements:} GHS mandate strict Service Level Agreements (SLAs) for their operations. The main requirements revolve around application availability (\% over time) and application response time (seconds). Moreover, GHS distinguishes between 2 tiers of service categories: the 1st tier represents time-critical services, which comprise all business processes involving medical transactions (inpatient/outpatient visits, prescribing, dispensing, lab test ordering, etc.); the 2nd tier concerns less time-critical business processes (e.g., enrolment or claims management). Availability percentages consider all planned downtimes (e.g., during deployments, patching or other maintenance work).

\noindent \textbf{Key Threats:} Healthcare systems have registered a significant uptick in cyber attacks, especially during the COVID-19 pandemic, with the top threats faced by the sector being malware and malicious insider activity. Data breaches may also trigger outages to the system until operations are deemed secure again, as was the case when hackers gained access to the records of 1.4M people who took the COVID-19 test in the Paris region in 2020\footnote{https://www.rfi.fr/en/france/20210916-hackers-steal-covid-test-data-of-1-4-million-people-from-paris-hospital-system}. Fines can be imposed to hospitals for the GDPR infringements (e.g., Centro Hospitalar Barreiro Montijo, 2018\footnote{https://blog.chino.io/gdpr-fines-in-helthcare-7-lessons/}), while ransomware attacks can also lead to devastating consequences (e.g., Düsseldorf University Hospital, Germany, 2020, 30 servers were held to ransom lead to the loss of a patient\footnote{https://dserver.bundestag.de/btd/19/242/1924247.pdf}). Considering the above, key threats include: Denial of Service attacks; Breaches of medical records; Back-office ransomware attacks. Prominent attack scenarios also include: Distributed Denial of Service Attacks to the GHS public web portal; Electronic Patient Records Data Breaches;

\section{Conclusions \& next steps} \label{sec:conclusion}

This paper presented the overall vision and main concepts behind PHOENI2X, aiming to design, deliver and validate an AI-enhanced CRF, providing orchestration, automation and response capabilities for business continuity and recovery, IR, and information exchange. Particular emphasis is given to aligning with the EU landscape, supporting cross-border, cross-organisational, and cross-functional collaboration and coordination tailored to the needs of OES, entities entrusted with the cybersecurity of MS, and other relevant EU cybersecurity bodies.

In terms of next steps, the individual components comprising PHOENI2X will be developed and integrated. Further, the resulting integrated framework will be validated in the context of 3 Essential Service use cases (Energy, Transport, Health) involving two OES, a provider in the supply chain of an OES, a telecom operator and two National cybersecurity Authorities. This will allow for a holistic assessment of PHOENI2X and its vision, collecting concrete evidence of its applicability, along with valuable feedback and pointers for its further refinement. 


\section*{Acknowledgment}
This work has received funding from the European Union’s Horizon Europe programme under Grant Agreement No. 101070586 (PHOENI2X project).

\bibliographystyle{IEEEtran}
\bibliography{biblio}

\end{document}